\documentclass[12pt]{article}
\usepackage{amsmath}
\usepackage{graphicx}
\usepackage{booktabs}
\usepackage{amsfonts}
\usepackage{hyperref}
\usepackage{lmodern}
\usepackage{amssymb}
\usepackage{tikz}
\usetikzlibrary{patterns}
\usetikzlibrary{snakes}

\newcommand{\be}{\begin{equation}}
\newcommand{\ee}{\end{equation}}

\newcommand{\bea}{\begin{eqnarray}}
\newcommand{\eea}{\end{eqnarray}}
\newcommand{\beqa}{\begin{eqnarray}}
\newcommand{\eeqa}{\end{eqnarray}}

\newcommand{\nn}{\nonumber}

\def\CD {{\cal D}}

\def\CF {{\cal F}}

\def\CM {{\cal M}}

\begin{document}

%\setlength{\baselineskip}{7mm}
%\begin{titlepage}

 \begin{flushright}
{\tt NRCPS-HE-02-2020}
\end{flushright}
  
\begin{center}
{\Large ~\\{\it   Yang-Mills Classical and Quantum Mechanics \\ and \\
Maximally Chaotic Dynamical Systems   
\vspace{1cm}

}

}%title ends

\vspace{1cm}
%\author{

%\author{
{\sl George Savvidy

\bigskip
\centerline{${}$ \sl Institute of Nuclear and Particle Physics}
\centerline{${}$ \sl NCSR Demokritos, Ag. Paraskevi,  Athens, Greece}
\bigskip

}%author ends
%}
%\date{}%in order NOT to write the date
%\maketitle
\end{center}
\vspace{20pt}

\centerline{{\bf Abstract}}
The maximally chaotic  dynamical systems (DS) are the systems which have nonzero Kolmogorov entropy. The  Anosov C-condition defines a reach class of hyperbolic dynamical systems that  have exponential  instability of the phase trajectories and positive Kolmogorov entropy and are therefore  maximally chaotic. The interest in Anosov-Kolmogorov systems is associated with the attempts to understand the relaxation phenomena, the foundation of the statistical mechanics, the appearance of turbulence in fluid dynamics, the non-linear dynamics of the Yang-Mills field, the N-body system in Newtonian  gravity and the relaxation phenomena in stellar systems and the Black hole thermodynamics. The classical- and quantum-mechanical properties of maximally chaotic dynamical systems,  the application of the C-K  theory to the investigation of the Yang-Mills dynamics and gravitational systems as well as their application in the Monte Carlo method will be presented.

%\vspace{12pt}

\noindent

%\end{titlepage}

 %\tableofcontents

%%%%%%%%%%%%%%%%%%%%%%%%%%%%%%%%%%%%

\pagestyle{plain}
 
\section{\it Introduction}

It seems natural to define the maximally chaotic dynamical K-systems as systems that have nonzero Kolmogorov entropy \cite{kolmo,kolmo1}. A large class of maximally chaotic dynamical systems was constructed by Anosov \cite{anosov}.  These are the systems that fulfil the C-condition. The Anosov C-condition leads to the exponential instability of phase trajectories, to the mixing of all orders and  positive Kolmogorov entropy. The examples of maximally chaotic systems were discussed in the earlier investigations   \cite{Artin,Hadamard, hedlund, hopf1,Hopf,anosov1,Gibbs,krilov,sinai3,turbul,kornfeld,arnoldavez} as well as in  \cite{rokhlin1,leonov,rokhlin2,smale,sinai2,sinai4,margulis,bowen0,bowen,gines}.  

In recent years the quantum-mechanical concept of maximally chaotic systems was developed in series of publications \cite{Shenker:2013pqa,Maldacena:2015waa,Gur-Ari:2015rcq,Cotler:2016fpe,Arefeva:1998,Arefeva:1998,Arefeva:1999,Arefeva:1999frh,Arefeva:2013uta} and references therein. It is based on the analysis of the black holes evaporation thermodynamics  and on the investigation of the so called out-of-time-order correlation functions conjectured to diverge exponentially with the exponent  linear in temperature and time  \cite{Shenker:2013pqa,Maldacena:2015waa,Gur-Ari:2015rcq,Cotler:2016fpe}.  The interrelation between the classical and quantum mechanical concepts of chaos will be discussed in the next sections.   

The article is organised as follows. In the second section  the  classification of the dynamical systems (DS)  by the increase of their statistical-chaotic properties will be presented   \cite{kornfeld,arnoldavez}.   These are ergodic, n-fold mixing and finally the K-systems, which have mixing of all orders and nonzero Kolmogorov entropy.   The question is: Do the maximally chaotic systems exist? 

The Anosov hyperbolic C-systems  represent a large class of  K-systems defined on the Riemannian manifolds of negative sectional curvatures and on high-dimensional tori. The general properties of the C-systems will be discussed in the third section. From the  C-condition follows the strong instability of the phase trajectories and, in fact, the instability is as strong as it can be in principle \cite{anosov,anosov1}. The distance between infinitesimally close trajectories increases exponentially and on a closed phase space of the dynamical system this leads to the uniform distribution of almost all trajectories over the whole phase space.    

The hyperbolic geodesic flow on closed Riemannian manifolds of negative sectional curvatures will be considered in the fourth section \cite{anosov,hedlund,hopf1,Hopf}.  It was proven by Anosov that manifolds with negative sectional curvatures fulfil the C-condition and therefore define a large class of maximally chaotic K-systems.  This result provides a powerful tool for the investigation of the Hamiltonian systems.    

In the fifth section the classical and quantum dynamics of the Yang-Mills  fields is reviewed \cite{Baseyan,Natalia,Asatrian:1982ga,SavvidyKsystem,Savvidy:1982jk,Savvidy:1984gi,Maldacena:2015waa,Gur-Ari:2015rcq,Arefeva:1998,Arefeva:1999,Arefeva:1999frh,Arefeva:2013uta,Chirikov,Nicolai}. In the case of space homogeneous gauge fields the Yang-Mills dynamics reduces to the classical-mechanical model, the Yang-Mills classical mechanics (YMCM) \cite{Baseyan,Natalia,Asatrian:1982ga,SavvidyKsystem,Savvidy:1982jk,Savvidy:1984gi}.  The sectional curvatures  are negative on the equipotential surface and generate exponential instability of the phase trajectories.  The question is to what extent the classical chaos influences the quantum-mechanical properties of the gauge fields. It is an example of fundamental quantum-mechanical matrix system,  the so called Yang-Mills quantum mechanics - YMQM \cite{ Savvidy:1982jk,Savvidy:1984gi}.     

The  other application of the C-K systems theory was found in the investigation of the relaxation phenomena in stellar systems  like globular clusters and galaxies \cite{body,Chandrasekhar,garry,Lang}.  The  Maupertuis's metric is used to reformulate the evolution of N-body system in Newtonian  gravity as a geodesic flow on a Riemannian manifold. Investigation of the sectional curvature allows to estimate the average value of the exponential divergency of the phase trajectories, {\it the correlation splitting time} $\tau_0$ and {\it the relaxation time} $\tau=\tau_0 \log(1/\delta v)$ \cite{yer1986a,Savvidy:2018ygo} in elliptic galaxies  and  globular clusters \cite{body}. This time is  shorter than the Chandrasekchar's binary relaxation time \cite{Chandrasekhar,Lang} and the Hubble time.   

Of special interest are continuous C-K systems which are defined on the two-dimensional surfaces embedded into the hyperbolic Lobachevsky plane \cite{Poghosyan:2018efd,Babujian:2018xoy,Savvidy:2018ffh,Gutzwiller}.  An example of such system has been defined in a brilliant article published in 1924 by the mathematician Emil Artin \cite{Artin}.   The differential geometry and group-theoretical methods \cite{Gelfand} are used to estimate  the correlation splitting time $\tau_0$  of  the classical correlation functions \cite{Poghosyan:2018efd}.

The quantum-mechanical properties of maximally chaotic systems are of special interest    \cite{ Savvidy:1982jk,Savvidy:1984gi,Maldacena:2015waa,Gur-Ari:2015rcq,Cotler:2016fpe,Poghosyan:2018efd,Babujian:2018xoy}.  In the eighth section the quantisation of the Artin system, the derivation of the  Maass wave function \cite{maass}, that describes the continuous spectrum, and the behaviour of the quantum-mechanical  correlation functions will be discussed \cite{Poghosyan:2018efd,Babujian:2018xoy}.  The spectral problem of quantum Artin system has deep number-theoretical origin  and was investigated in a series of pioneering articles \cite{maass,roeleke,selberg1,selberg2,bump,Faddeev,Faddeev1,hejhal2,hejhal}. 
The that two- and four-point correlation functions decay exponentially in time, reminiscent to the correlation splitting time $\tau_0$ of the system in the classical regime \cite{Poghosyan:2018efd}.  The commutator $C(\beta,t)$ grows exponentially  \cite{Babujian:2018xoy}, mimicking  the expansion of the initial phase space volume $\delta v$ over the whole phase space with an exponential rate $\tau=\tau_0 \log(1/\delta v)$ \cite{yer1986a}(10-12,17),\cite{Savvidy:2018ygo} .      
 
In the ninth section we shall demonstrate that the Riemann zeta function zeros \cite{Riemann} define the position and the widths of the resonances of the quantised Artin system  
\cite{Savvidy:2018ffh}.  A possible relation of the zeta function zeros and quantum-mechanical spectrum was discussed in the past:  David Hilbert seems to have proposed the idea of finding a system whose spectrum contains the zeros of the Riemann $\zeta(s)$ function \cite{Gutzwiller}.  The  poles of the S-matrix are located in the  complex plane  and are expressed in terms of zeros $u_n$ of the Riemann zeta function  $\zeta(\frac{1}{2} - i u_n) =0,~n=1,2,....$  
\be
E = E_n - i {\Gamma_n \over 2}= ({u^2_n \over 4} + {3\over 16}) - i {u_n \over 2}, \nn
\ee
where $E_n$ is the energy and $\Gamma_n$ is the width of the n'th resonance  \cite{Savvidy:2018ffh}. 

In the tenth section the attention will be turn to the investigation of the C-K systems that are defined on high-dimensional tori \cite{anosov}.  It was suggested in 1986 \cite{yer1986a} to use the C-K systems to generate high quality pseudorandom numbers for Monte-Carlo simulations   
\cite{konstantin,Savvidy:2015jva,Savvidy:2015ida,Savvidy:2018ygo,hepforge}. The high entropy MIXMAX generator based on C-K system was implemented into the Geant4/CLHEP and ROOT scientific toolkits at CERN \cite{hepforge,geant,cern,root}.

\section{\it  Hierarchy of Dynamical Systems. Kolmogorov Entropy}

In ergodic theory the dynamical systems (DS) are classified by the increase of their statistical-chaotic properties  \cite{kornfeld,arnoldavez}.  Let $x=(q_1,...,q_d,p_1,...,p_d) $ be a point of the phase space $x \in M $ of the Hamiltonian systems that is equipped with a positive Liouville measure $d\mu(x)= \rho(q,p) dq_1...dq_d dp_1...dp_d$, which is invariant under the Hamiltonian flow.  The operator $T^t x = x_t$ defines the time evolution of the trajectories.   The n-fold mixing takes place if for any number of  sets  $A_1,...,A_n \subset  M$ the  
$
\lim_{t_n,...,t_1 \rightarrow \infty} \mu[ T^{t_n} A_n \cap ....T^{t_2} A_2 \cap T^{t_1} A_1 \cap  B ]  
= \mu[A_n]...   \mu[A_2] \mu[A_1] \mu[B]
$
or alternatively \cite{Gibbs,kornfeld}
\be\label{mixnn}
\CD_{t}(f_n,...,f_1) =   
   \lim_{t_n,...,t_1 \rightarrow \infty}  \langle f_n(T^{t_n}x).....f_1(T^{t_1}x)  \rangle - \langle f_n(x)\rangle .....\langle f_1(x)\rangle   =0.
\ee
A  class of dynamical systems which have even stronger chaotic properties was introduced by Kolmogorov in \cite{kolmo,kolmo1}.  These are the DS that have a non-zero  entropy, so called  quasi-regular DS, or   K-systems.  Let $\alpha = \{A_i\}_{i \in I}$ ( $I$ is finite or countable)  be a  measurable partition of 
the phase space $M$ into the non-intersecting subsets $A_i$ which cover the whole phase space $M$
and then define the entropy of the partition $\alpha$ as 
$
h(\alpha) = - \sum_{i \in I} \mu(A_i) \ln \mu(A_i).
$
The entropy of the partition $\alpha$ with respect to the discrete evolution $T^n$ 
is defined as a limit
$
h(\alpha, T)= \lim_{n \rightarrow \infty} {h(\alpha \vee T \alpha \vee ...\vee T^{n-1} \alpha) \over n},~
n=1,2,...
$ \cite{kolmo,kolmo1,sinai3,rokhlin1,rokhlin2}.
This number is equal to the entropy of the partition  
$
\beta = \alpha \vee T \alpha \vee ...\vee T^{n-1} \alpha 
$
generated  by the iteration of the partition $\alpha$ 
by the  evolution operator $T$. The entropy of   
  $T$ is defined as a supremum: 
$
h(T) = \sup_{\{ \alpha \}} h(\alpha,T),
$
taken over all partitions $\{ \alpha \}$ of  $M$. 
It was proven that the K-systems have mixing of all orders: K-mixing $\supset$ infinite mixing, $\supset$,n-fold mixing,$\supset$ mixing $\supset$ ergodicity \cite{kolmo,kolmo1,sinai3,rokhlin1,rokhlin2,sinai4,gines}.  The question is: Do maximally chaotic systems exist? The Anosov hyperbolic C-systems \cite{anosov} represent a large class of  K-systems considered  in the next section. 

\section{\it  Hyperbolic Anosov C-systems} 

In the fundamental work on geodesic flows on closed
Riemannian manifolds $Q^n$ of negative sectional curvatures \cite{anosov} Anosov
pointed out  that the basic property of the geodesic flow on such manifolds 
is the  uniform exponential  instability of phase trajectories.
The exponential instability of geodesics was studied by  Lobachevsky, Hadamard, Artin \cite{Artin},  Hedlund  \cite{hedlund} and Hopf \cite{Hopf} and others.
The concept of exponential instability appears to be extremely rich, and Anosov suggested to elevate it into a fundamental property of a new class of dynamical systems
which he called C-systems\footnote{The letter C is used because these systems 
fulfil the "C-condition" (\ref{ccondition})\cite{anosov}. }. 
The brilliant idea to consider dynamical systems 
which have  local and homogeneous  hyperbolic instability of phase trajectories
is appealing to the intuition and  has  very deep  physical content.   
The richness of the concept is expressed by the fact  that C-systems 
occupy a nonzero volume  in the space of dynamical 
systems \cite{anosov}.  Anosov provided an extended list  of   C-K systems, these are:  {\it i)   C-cascades  and  ii) the geodesic flow on the Riemannian manifolds of  variable negative sectional curvatures} \cite{anosov}.   

\begin{figure}
 \centering
\includegraphics[width=6cm]{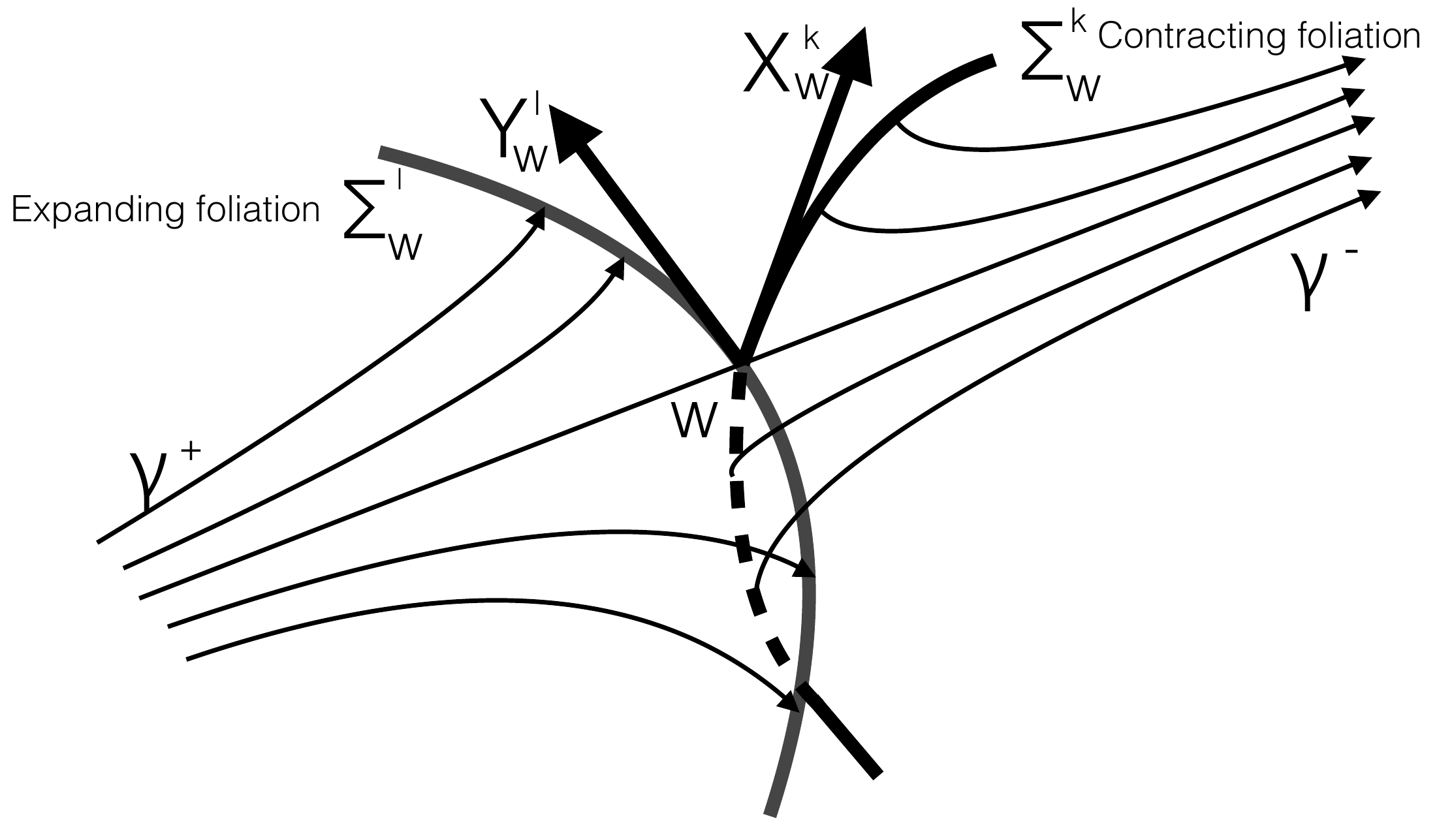}
\centering
\caption{ At each point $w$ of the C-system the tangent space $R^{m}_{w}$  
is  decomposable  into a direct sum of two linear spaces  $Y^l_{w} $ and $X^k_{w} $. 
The expanding and contracting geodesic flows
are $\gamma^+$ and $\gamma^-$. The expanding and 
contracting invariant  foliations  $\Sigma^l_{w} $ and $\Sigma^k_{w} $ 
 are transversal to the geodesic flows and their corresponding tangent spaces 
are  $Y^l_{w} $ and $X^k_{w} $.  } 
\label{fig13}
\end{figure}
The C {\it cascade}  on a d-dimensional compact phase space $M^{d}$ is induced by 
the diffeomorphisms $T: M^d \rightarrow M^d$ \cite{anosov}.  The iterations
are defined by  a repeated action of the operator  
$\{ T^n, -\infty < n < +\infty  \}$, where $n$ is an  integer number.  
The tangent space at 
the point $x \in M^d$ is  denoted by $R^d_{x}$. 
The C-condition requires that the tangent space $R^{d}_{x}$ at each point $x$
of the d-dimensional phase space $M^{d}$ of the dynamical system $\{T^{n}\}$ 
should be decomposable  into a 
direct sum  $ R^{d}_{x}= X^{k}_{x} \bigoplus Y^{l}_{x}$   of  two linear spaces  $X^{k}_{x}$ and $Y^{l}_{x}$ with the following 
properties \cite{anosov}:
\bea\label{ccondition}
 &~~a) \vert \tilde{T}^{n} \xi  \vert  \leq ~ a \vert   \xi \vert e^{-c n}~ for ~n \geq 0 ; ~
\vert \tilde{T}^{n} \xi  \vert  \geq~ b \vert \xi \vert e^{-c n} ~for~ n \leq 0,~~~\xi \in  X^{k}_{x}, \nn\\
&b) \vert \tilde{T}^{n} \eta  \vert  \geq~ b \vert \eta \vert e^{c n} ~~for~ n \geq 0;~
\vert \tilde{T}^{n} \eta  \vert  \leq ~ a \vert   \eta \vert e^{c n}~ for~ n \leq 0,~~~\eta \in Y^{l}_{x},
\eea
where the constants a,b and c are positive and are the same for all $x \in M^d$ and all 
$\xi \in  X^{k}_{x}$, $\eta \in Y^{l}_{x}$. The linear spaces $X^{k}_{x}$ and $Y^{l}_{x}$ are invariant  with respect to the derivative  mapping  $\tilde{T}^{n} X^{k}_{x} = X^{k}_{T^n x}, ~
\tilde{T}^{n} Y^{l}_{x} = Y^{l}_{T^n x}$ and represent the exponentially {\it contracting and expanding 
linear spaces} (see Fig.\ref{fig13}). The    $X^{k}_{x}$ and $Y^{l}_{x}$ are the target vector spaces  to 
the contracting and expanding foliations  $\Sigma^k_x$ and $\Sigma^l_x$   \cite{anosov}.  {\it The C-condition represents a powerful criterion for the identification of maximally chaotic DS and the main task reduces to the examination of the C-condition in each particular case}.

\section{\it The Hyperbolic C-systems on  Riemannian Manifolds }

Let us consider a  Riemannian manifold $Q$ with the local coordinates $q^{\alpha}(s) \in Q$, $\alpha=1,2,....,3 N$ and the velocity vector
$
u^{\alpha}= {d q^{\alpha} \over d s},~\alpha=1,2,....,3 N.
$
The proper time   $s$ along the $ q^{\alpha}(s)$ is equal to  the length, while   
the Riemannian metric on $Q$ is defined as 
$ds^2 = g_{\alpha\beta} d q^{\alpha} d q^{\beta}$,
and therefore $
g_{\alpha\beta}u^{\alpha} u^{\beta}=1.
$
The infinitesimal deformation of  geodesics  congruence $ q^{\alpha}(s,\upsilon)$ is defined as a  vector 
$
\delta q^{\alpha} = {\partial q^{\alpha} \over \partial \upsilon} d \upsilon.
$
The  resulting phase space manifold $(q(s),u(s)) \in M$ has a bundle structure 
with the base $q \in Q $ and the  spheres 
$S^{3N-1}$ of unit tangent vectors $u^{\alpha}$  as fibers.
The geodesic equation has the form
$
{d^2 q^{\alpha} \over d s^{2}}+ \Gamma^{\alpha}_{\beta\gamma} {d  q^{\beta} \over d s }
{d  q^{\gamma} \over d s }=0
$,
and the Jacobi equation for the relative acceleration depends on the Riemann curvature:
$
{D^2 \delta q^{\alpha} \over ds^2}
 = - R^{\alpha}_{\beta\gamma\sigma} u^{\beta} \delta q^{\gamma} u^{\sigma} .
$
Following Anosov it is convenient to define the deviation norm 
$
 \vert \delta q  \vert^2   \equiv g_{\alpha\beta} \delta q^{\alpha} \delta q^{\beta}
$
and  represent the Jacobi equation  in the form
 \bea
 {d^2\over ds^2} \vert \delta q  \vert^2 
 =-2 K(q,u,\delta q) ~\vert u \wedge \delta q \vert^2 +2 \vert  \delta u  \vert^2 ,~~~~~~~
\eea
where $K(q,u,\delta q)$ is the sectional curvature  in the two-dimensional directions 
defined by the velocity vector 
$u^{\alpha}$ and the deviation vector $\delta q^{\beta}:$
$
 K(q,u, \delta q)  = { R_{\alpha\beta\gamma\sigma} \delta q^{\alpha} 
 u^{\beta} \delta q^{\gamma}  u^{\sigma} \over  \vert u \wedge \delta q  \vert^2 } . 
$
In the decomposition     
 $
 \delta q  = \delta q _{\perp} + \delta q_{\parallel},
 $ 
the $\delta q_{\parallel}$  describes a translation 
along the geodesic trajectory and has no physical interest, the transversal 
component $\delta q_{\perp}$ describes a physically relevant distance between original and infinitesimally close  trajectories $\vert u \delta q_{\perp} \vert =0$.  
For the  transversal deviation one can get \footnote{The area spanned 
by the bivector is simplified:  $\vert u \wedge \delta q_{\perp} \vert^2= \vert u \vert^2  \vert   \delta q_{\perp}  \vert^2 - \vert u \delta q_{\perp} \vert^2 = \vert \delta q_{\perp} \vert^2$.}:
$
{d^2\over ds^2} \vert \delta q_{\perp}  \vert^2 = 
-2 K(q,u,\delta q_{\perp}) ~\vert \delta q_{\perp} \vert^2 
+2 \vert   \delta u_{\perp}  \vert^2. 
$
The last term is positive definite, therefore for the 
{\it relative acceleration}: 
$
{d^2\over ds^2} \vert \delta q_{\perp}  \vert^2 \geq
-2 K(q,u,\delta q_{\perp}) ~\vert \delta q_{\perp} \vert^2. 
$
If the sectional curvatures are negative and uniformly bounded from above by a constant $\kappa$:
$
K(q,u,\delta q_{\perp}) \leq - \kappa < 0, ~\text{where}~ \kappa = \min_{(q, u,\delta q_{\perp})} \vert K(q,u,\delta q_{\perp}) \vert 
$, 
then \cite{anosov}
\bea\label{anosovinequality1}
{d^2\over ds^2} \vert \delta q_{\perp}  \vert^2 \geq
2 \kappa ~\vert \delta q_{\perp} \vert^2.
\eea
The space of solutions  is divided into two linear spaces  $X_q$ and $Y_q$. The space $Y_q$ consists of the solutions with positive first derivative 
$
{d \over ds} \vert \delta q_{\perp}(0)  \vert^2 > 0    
$
and exponentially expanded: 
\be\label{grows}
\vert \delta q_{\perp}(s) \vert \geq {1\over 2} \vert \delta q_{\perp}(0) \vert e^{\sqrt{2\kappa} s},
\ee
while the space $X_q$ consists of the solutions  with negative first derivative
$
  {d \over ds} \vert \delta q_{\perp}(0)  \vert^2  < 0 
  $ 
and exponentially contracted:  
\be\label{decay}
\vert \delta q_{\perp}(s) \vert \leq {1\over 2} \vert \delta q_{\perp}(0) \vert e^{-\sqrt{2\kappa} s}.
\ee
This proves that the geodesic  flow (\ref{anosovinequality1}) fulfils the C-conditions (\ref{ccondition}).  The system is maximally chaotic  and tends to equilibrium with exponential rate.  The correlation splitting time is: 
$
\tau_0 = 1/\sqrt{2\kappa}
$
\cite{yer1986a,Savvidy:2018ygo}.
The Anosov theory of hyperbolic C-K systems found application in the investigation of gauge  and gravitational systems  \cite{Baseyan,Natalia,Asatrian:1982ga,SavvidyKsystem,Savvidy:1982jk,Savvidy:1984gi,body}.   

\section{\it The Yang-Mills Classical and Quantum Mechanics}

In the case of space-homogeneous gauge fields $\partial_i A^a_k =0$, $i,k=1,...,d$ the Yang-Mills system reduces to the Hamiltonian system of the form \cite{Baseyan,Natalia,Asatrian:1982ga,SavvidyKsystem,Savvidy:1982jk,Savvidy:1984gi}
\be\label{YMclassical}
H= \sum_{i} {1\over 2} Tr \dot{A}_{i} \dot{A}_{i} + {g^2\over 4} \sum_{i,j}Tr [A_i,A_j]^2,
\ee
where the index $a=1,...,N^2-1$ for $SU(N)$ group and the constraint has the form
$
\hat{n}=[\dot{A}_{i}, A_i]=0.
$
It is natural to call this system the Yang-Mills Classical Mechanics (YMCM) \cite{SavvidyKsystem,Savvidy:1982jk,Savvidy:1984gi}. It is a mechanical system with $d \cdot(N^2-1)$ degrees of freedom.  The YMCM has a number of conserved integrals: the space and isospin angular momenta  
and the energy integral (\ref{YMclassical}). The question is if there exist additional conservation integrals. Performing the substitution 
$
A = O_1 E O^T_2,
$
when  $d=3$, $N=2$ and  $E=(x(t),y(t),z(t))$ is a diagonal matrix and $O_1,O_2$ are orthogonal matrices the Hamiltonian (\ref{YMclassical}) will take the form \cite{Natalia,Asatrian:1982ga,SavvidyKsystem,Savvidy:1982jk}
\be\label{YMCM}
H_{FS}= {1\over 2}  (\dot{x}^2 + \dot{y}^2 + \dot{z}^2 )+  {g^2\over 2}( x^2y^2 +y^2z^2 +z^2x^2 ) + T_{YM},
\ee
where $T_{YM}$ is  the energy of the Yang-Mills "top" spinning in space and isospace.  The YMCM is equivalent to the geodesic flow on a Riemannian manifold with the Maupertuis's metric and the sectional curvatures on the equipotential surface $x^2y^2 +y^2z^2 +z^2x^2=1$ are negative and generate the exponential instability of the trajectories.   The  question is: What are the quantum-mechanical properties of the classically chaotic system  (\ref{YMclassical}), (\ref{YMCM}) and what is the structure of its spectrum and of the wave functions?  The Schr\"odinger equation of the Yang-Mills quantum-mechanical system (YMQM) represents a special class of quantum-mechanical matrix  models \cite{ Savvidy:1982jk,Savvidy:1984gi}.   The ground state wave function $\Psi(x) = \Phi(x)/\sqrt{D(x)}$ fulfils the equation \cite{Savvidy:1984gi}
\beqa\label{groundwave}
 -{1\over 2} \partial^2_i  \Phi +  {1\over 2} \sum_{i<j} \Big( {1\over (x_i-x_j)^2} +{1\over (x_i+x_j)^2}   + g^2  x^2_i x^2_j\Big) \Phi=   E \Phi ,
\eeqa
where $D(x) = \vert (x^2_1 - x^2_2)(x^2_2 - x^2_3)(x^2_3 - x^2_1)\vert$.  
The analytical investigation of the equation (\ref{groundwave}) is a challenging problem because it  cannot be solved by separation of variables as far as all canonical symmetries are already extracted and the residual system possesses no continuous symmetries. The energy spectrum of YMQM system is discrete  \cite{ Savvidy:1982jk,Savvidy:1984gi}, the energy levels "repulse" similar to the eigenvalues of the matrices with randomly distributed elements \cite{Wigner,Mehta,Dyson}. In the next section   the gravitational $N$-body problem will be presented \cite{body}. 

\section{\it Collective Relaxation of Stellar Systems}

The $N$-body problem in Newtonian gravity can be formulated as a geodesic flow on Riemannian manifold with the conformal Maupertuis  metric  
$
ds^2 = (E-U) d \rho^2 = W \sum^{3N}_{\alpha=1} (d q^{\alpha})^2,
$
where $W=E-U$, the $\{q^{\alpha}\}$ are the coordinates of the stars
$
\{q^{\alpha}\} = \{  M^{1/2}_1 \vec{r}_1,.....,M^{1/2}_N \vec{r}_N      \},~\alpha =1,...,3N
$ 
and $U=-G \sum_{a < b}{M_a M_b \over \vert \vec{r}_a - \vec{r}_a \vert }$.
The sectional curvature has the form \cite{body}
\bea\label{sectional2}
R_{\alpha\beta\gamma\sigma} \delta q^{\alpha}_{\perp} 
 u^{\beta} \delta q^{\gamma}_{\perp}  u^{\sigma}= \Big[-{1\over 3N} {\triangle W \over  W^2} -({1\over 4} - {1\over 2N}){(\nabla W)^2 \over W^3}\Big]  \vert \delta q_{\perp}   \vert^2.
\eea
The number of stars in a galaxy is very large, $N \gg 1$ and the dominant term in sectional curvature (\ref{sectional2}) is negative: 
$
K(q,u, \delta q)  =   - {1\over 4}  {(\nabla W)^2 \over W^3}  < 0.
$
The deviation equation (\ref{anosovinequality1})  will take the form 
$
{d^2\over dt^2} \vert \delta q_{\perp}  \vert^2 \geq
  {(\nabla W)^2 \over W}  ~\vert \delta q_{\perp} \vert^2
$
and the correlation splitting time $\tau_0$ can be defined as 
$
\tau_0 = \sqrt{{W \over (\nabla W)^2 }}
$
\cite{yer1986a,Savvidy:2018ygo,body}.
For the elliptic galaxies and globular clusters it is
\be
\tau_0 \approx 10^8 yr \Big({v \over 10 km/s}\Big) \Big( {n \over 1 pc^{-3}}\Big)^{-2/3} \Big( { M \over M_{\odot}  }\Big)^{-1}. 
\ee
The relaxation  time $\tau = \tau_0 \log(1/\delta v)$ is by few orders of magnitude shorter than the Chandrasekchar's binary relaxation time \cite{Chandrasekhar,garry,Lang} and is less than the Hubble time.  

\section{\it  Correlation Functions of Classical  Artin  System}

\begin{figure}[htbp]
	 \hspace*{-4cm} 
	\begin{tikzpicture}[scale=1.5]
	\clip (-4.3,-0.4) rectangle (4.5,2.8);
	\draw[,->] (-2.1,0) -- (2.1,0)  node[anchor=north west] {$x$ };
	\draw[dashed] (0,0) -- (0,1.2);
	\draw[dashed,->]  (0,1.8)--(0,2.2)  node[anchor=south east] {$y$};
	\foreach \x in {-1,-0.5 ,0,0.5,1}
	\draw (\x cm,1pt) -- (\x cm,-1pt) node[anchor=north] {$\x$};
	\foreach \y in {}
	\draw (1pt,\y cm) -- (-1pt,\y cm) node[anchor=east] {$\y$};
	\draw (1,0) arc (0:180:1cm);
	\draw (0.5,0)--(0.5,0.86602540378);
	\draw (-0.5,0)--(-0.5,0.86602540378);
	\draw[ ultra thick, black] (0.5,0.86602540378)--(0.5,2.1);
	\draw[ ultra thick, black] (-0.5,0.86602540378)--(-0.5,2.1);
	\draw[ultra thick, black] (0.5,0.86602540378) arc (60:120:1cm);
	\draw (-0.6,0.7) node{ $A$};
	\draw (0.4,0.7) node{ $B$};
	\draw (-0.1,0.85) node{ $C$};
	\draw [->,snake=snake,
	segment amplitude=.6mm,
	segment length=1.5mm,
	line after snake=1mm] (-0.3,2) -- (-0.3,1.3);
	\draw [->,snake=snake,
	segment amplitude=.6mm,
	segment length=1.5mm,
	line after snake=1mm] (0.3,1.3)-- (0.3,2);
	%%%%%%%%%%%%%%%%%%%%%%%%%%%%%%%%%%%%%%%%%%%%%%%%%%%%
	\draw [->,snake=snake,
	segment amplitude=.6mm,
	segment length=1.5mm,
	line after snake=1mm] (0.1,1.5) -- (0.2,1.7);
	%%%%%%%%%%%%%%%%%%%%%%%%%%%%%%%%%%%%%%%%%%%%%%%%%%%
	\draw [->,snake=snake,
	segment amplitude=.6mm,
	segment length=1.5mm,
	line after snake=1mm] (-0.1,1.5) -- (-0.2,1.7);
	%%%%%%%%%%%%%%%%%%%%%%%%%%%%%%%%%%%%%%%%%%%%%%%%%%%
	%%%%%%%%%%%%%%%%%%%%%%%%%%%%%%%%%%%%%%%%%%%%%%%%%%%
	\draw [->,snake=snake,
	segment amplitude=.6mm,
	segment length=1.5mm,
	line after snake=1mm] (-0.1,1.4) -- (-0.2,1.2);
	%%%%%%%%%%%%%%%%%%%%%%%%%%%%%%%%%%%%%%%%%%%%%%%%%%%
	%%%%%%%%%%%%%%%%%%%%%%%%%%%%%%%%%%%%%%%%%%%%%%%%%%%
	\draw [->,snake=snake,
	segment amplitude=.6mm,
	segment length=1.5mm,
	line after snake=1mm] (0.1,1.4) -- (0.2,1.2);
	%%%%%%%%%%%%%%%%%%%%%%%%%%%%%%%%%%%%%%%%%%%%%%%%%%%
	\draw (1.2,1.8) node{ $\frac{\theta({1\over 2}+i p)}{\theta({1\over 2}-i p)}e^{i p \tilde{y}}$};
	\draw (-1,1.8) node{ $e^{-i p \tilde{y}}$};
	%%%%%%%%%%%%%%%%%%%%%%%%%%%%%%%%%%%%%%%%%%%%%%%%%%%%%%%
	\draw (0.08,2.5) node{ $\CD$};
	\end{tikzpicture}
	 \includegraphics[angle=0,width=3cm]{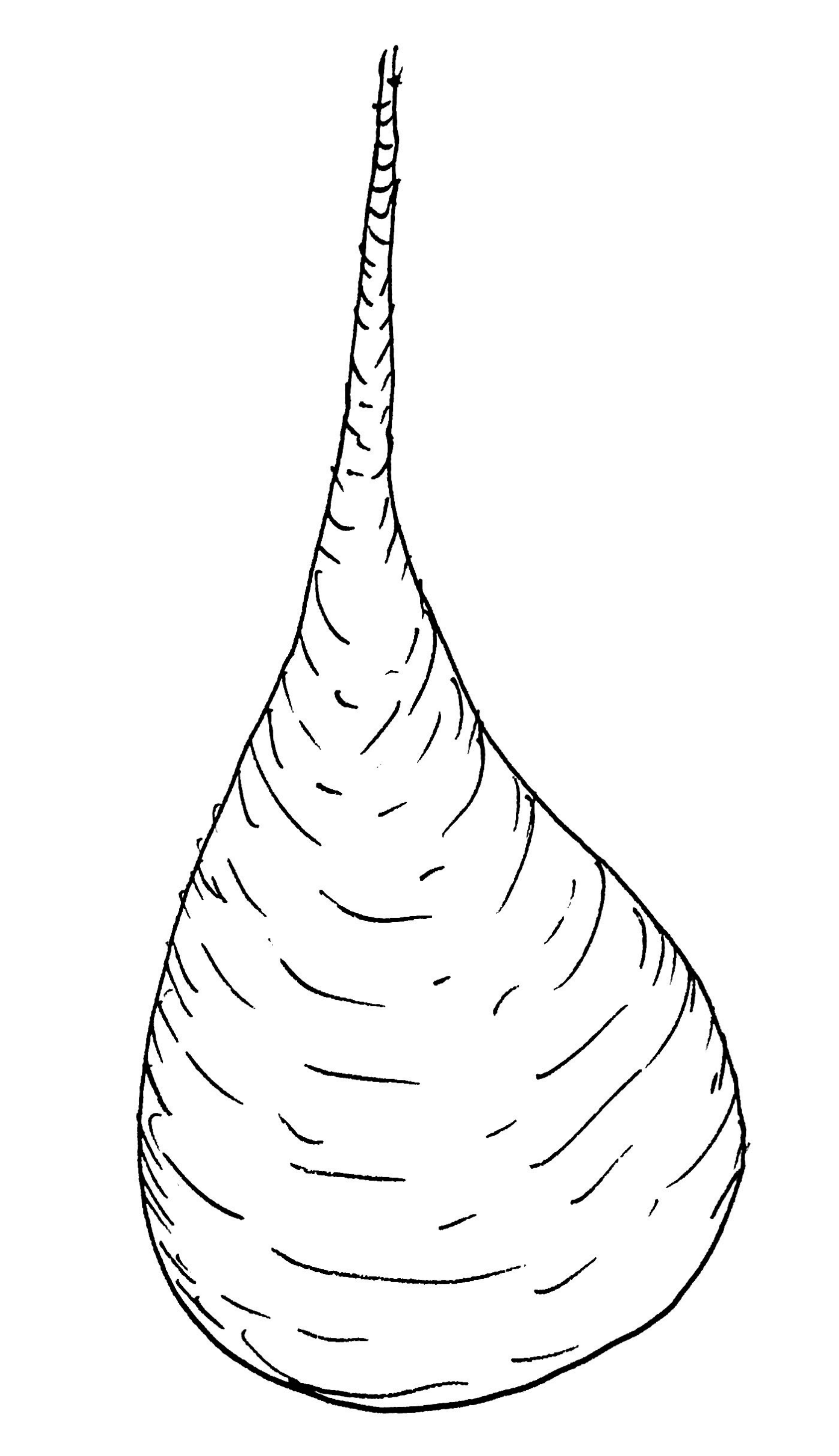}

\caption{The  fundamental region $\CF$ is the hyperbolic triangle $AB\CD$. The edges are the arc $AB$ and the rays $AD$, $BD$. The edges are  identified and form the Artin "bottle"  of  a non-compact surface $ \bar{\CF}$ of a constant negative curvature and sphere topology with a cusp on the north pole shown on the right figure.  The  Maass plane waves are given by formula (\ref{alterwave0}). The Gauss-Bonne integral is  $\int K \sqrt{g} d^2 \xi = (-1) {\pi \over 3} + (2\pi - 2{ \pi \over 3})+ (2\pi - 2 { \pi \over 2}) + (2\pi - 0 ) =4\pi$.}
	
	\label{scattering}
\end{figure}
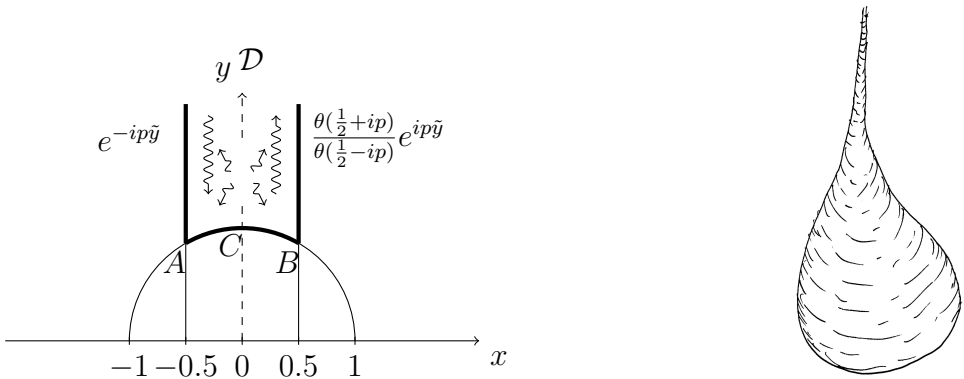

The Artin system \cite{Artin} is defined on the fundamental region $\CF$ of the Lobachevsky plane obtained by the identification of points congruent with respect to the modular group $SL(2,Z)$, a discrete subgroup of the Lobachevsky plane isometries $SL(2,R)$.  The fundamental region $\CF$ in this case is a hyperbolic triangle \cite{Poghosyan:2018efd,Babujian:2018xoy,Savvidy:2018ffh} shown on Fig.\ref{scattering}.  The time evolution of the physical observables $\{f(x,y, \theta)\}$ is defined on the phase space  $(x,y,\theta) \in \CM$, where $\tau =x+iy \in \CF$ and $\theta \in S^1$ is a
direction of a unit velocity vector.   The Liouville measure is  
$
d \mu = \frac{dx dy}{y^2} d\theta
$
and the invariant product of functions  is  
$
(f_1 ,f_2)
= \int_{0}^{2\pi} d \theta \int_{\CF}f_1(\theta,x,y) \overline{f_2(\theta,x,y) }
{d x d y \over y^2}
$ 
\cite{Poincare,Gelfand}.
A  two-point correlation function (\ref{mixnn}) has the form \cite{Poghosyan:2018efd}:
\beqa
\CD_{t}(f_1,f_2) &=&    
= \int_{0}^{2\pi}\int_{\CF}f_1[x,y,\theta] ~\overline{f_2[x'(x,y,\theta,t), y'(x,y,\theta,t), \theta'(\theta,t)] }
{d x d y \over y^2}d \theta.
\eeqa
We found that the upper bound on the correlations functions is \cite{Poghosyan:2018efd}
\beqa\label{splittingtime}
\vert \CD_t(f_1,f_2) \vert  & \leq &     ~ C_{f_1 f_2}~  e^{-{N\over 2} K \vert t \vert},~
\eeqa
where the Poincar\'e  metric is $ds^2 = {dx^2 +dy^2 \over K  y^2}
$   and $N$ is the weight of the automorphic functions. {\it The correlation splitting time} is therefore  
$
\tau_0 = {2\over N K }.
$
Considering a set of initial trajectories occupying a small volume $\delta v$  in the phase space of a C-K system, one can ask how fast this small phase volume will be uniformly distributed over the whole phase space \cite{yer1986a,Savvidy:2018ygo}. {\it This characteristic time interval $\tau$ defines the relaxation time at which the system reaches a stationary distribution} \cite{yer1986a,Savvidy:2018ygo}. Because the entropy defines the expansion rate of the phase space volume it follows that $\tau =\tau_0  \log(1/ \delta v)$ with large  hierarchy between $\tau_0$ and  $\tau$ \cite{yer1986a,Savvidy:2018ygo}.

\section{ \it Correlation Functions of Quantum   Artin  System } 
 
As we have seen, the classical correlation functions decay exponentially  \cite{Poghosyan:2018efd,Savvidy:2018ygo}, and our aim is to investigate the behaviour of the quantum-mechanical correlation functions  \cite{Babujian:2018xoy,Savvidy:2018ffh}.  It was conjectured  in \cite{Maldacena:2015waa} that the behaviour of the out-of-time-order correlation functions play a  crucial role in diagnosing the classical chaos. To investigate the correlation functions it is necessary  to know the spectrum and the corresponding wave functions.   The spectral problem has deep number-theoretical origin    
\cite{maass,roeleke,selberg1,selberg2,bump,Faddeev,Faddeev1,hejhal2,hejhal}. The continuous spectrum is described by 
the Maass wave function \cite{maass} and has the form \cite{Babujian:2018xoy,Savvidy:2018ffh}
\bea\label{alterwave0}
&\psi_{p} (x,\tilde{y})  
= e^{-i p \tilde{y}  }+{\theta(\frac{1}{2} +i p) \over \theta(\frac{1}{2} -i p)}   \, e^{  i p \tilde{y}}  +{ 4  \over  \theta(\frac{1}{2} -i p)}   \sum_{l=1}^{\infty}\tau_{i p}(l)
K_{i p }(2 \pi  l e^{\tilde{y}} )\cos(2\pi l x),
\eea
where $\int dy/y =  \ln y = \tilde{y} $ is a  physical  distance on the fundamental triangle  and the corresponding momentum is $p_y$.   The $e^{-i p \tilde{y}  }$ describes the incoming  and $e^{  i p \tilde{y}}$ the outgoing plane waves, the reflection amplitude is a pure phase 
$
S(p) = {\theta(\frac{1}{2} +i p) \over \theta(\frac{1}{2} -i p)} = \exp{[i\, \varphi(p)]},
$
where
$ 
\theta(\frac{1}{2} -i p)=  {\zeta (1- 2 i p) \Gamma (\frac{1}{2} -i p) \over  \pi ^{\frac{1}{2} -i p} }.
$
The continuous energy spectrum is given by the formula  
$
E=   p^2  + \frac{1}{4} .
$
The wave functions of the discrete spectrum  were known only numerically \cite{selberg1,selberg2,hejhal}. The correlation functions defined in \cite{Maldacena:2015waa} are: 
$
\CD_2(\beta,t)=  \langle   A(t)   B(0) e^{-\beta H}   \rangle$,~ 
$ \CD_4(\beta,t)=  \langle  A(t)   B(0) A(t)   B(0)e^{-\beta H}   \rangle$ and $C(\beta,t) =   -\langle [A(t),B(0)]^2 e^{-\beta H} \rangle
$. We were considering the  Louiville type operators $A$ and $B$ of the form:
$
A(N)=  e^{-2 N \tilde{y}},~N=1,2,.....
$
\cite{Babujian:2018xoy}. We found that two- and four-point correlation functions decay exponentially in time, reminiscent to the correlation splitting time $\tau_0$ of the system in the classical regime (\ref{splittingtime})\cite{Poghosyan:2018efd}.  The commutator $C(\beta,t)$ grows exponentially  \cite{Babujian:2018xoy}, mimicking  the expansion of the initial phase space volume $\delta v$ growing  at an exponential rate $\tau=\tau_0 \log(1/\delta v)$ \cite{yer1986a,Savvidy:2018ygo}.   

\section{\it  Resonances and Riemann Zeta Function Zeros  }

Here we shall demonstrate that the Riemann zeta function zeros define the position and the widths of the resonances of the quantised Artin system \cite{Savvidy:2018ffh}.  In physical terms  it is narrow infinitely long waveguide stretched out to infinity and a cavity resonator attached to it at the bottom, Fig.\ref{scattering}.   As  the energy of the incoming wave comes close to the eigenmodes of the cavity a pronounced resonance behaviour shows up in the scattering amplitude \cite{Savvidy:2018ffh}.  The  position of the S-matrix poles is defined by the absence of the incoming waves in (\ref{alterwave0})
$
 \theta(\frac{1}{2} -i p_n) =0,
$
\cite{Savvidy:2018ffh} and is determined by zeros  of the Riemann zeta function
$
\zeta(\frac{1}{2} - i u_n) =0,~ n=1,2,.... 
$ \cite{Riemann}. 
The location of poles is  at  complex momenta 
$
p_n = {u_n \over 2} - i \,{1 \over 4}, ~n=1,2,..... 
$,
thus:
\be\label{resonances34} 
E = p^2_n +{1 \over 4}~  = ~{u^2_n \over 4} + {3\over 16}
- i \, {u_n \over 4}.
\ee
These are the resonances 
$
E = E_n - i {\Gamma_n \over 2}, 
$
where
$E_n = {u^2_n \over 4} + {3\over 16}$ and $\Gamma_n = {u_n \over 2}$.

\section{\it  C-cascades and MIXMAX Random Number Generator}

Let us consider the second class of the C-K systems defined on high-dimensional tori \cite{anosov} and their application in the Monte Carlo method \cite{yer1986a,konstantin,Savvidy:2015jva,Savvidy:2015ida,Savvidy:2018ygo,hepforge}. The automorphisms of a torus are generated by the transformations  
$
x_i \rightarrow \sum^{n}_{j=1} T_{ij} x_j,~(mod ~1),
$
where   $Det~ T =1$ and  matrix $T$ has no eigenvalues on the unit circle.   The entropy of the Anosov automorphisms on a torus is equal to the sum \cite{anosov,smale,sinai2,margulis,bowen0,bowen}:
$
h(T) = \sum_{\vert \lambda_{\beta} \vert > 1} \ln \vert \lambda_{\beta} \vert
$
and depends on the spectrum  of the operator $T$.  The strong instability of phase trajectories  leads to the statistical behaviour of the Anosov C-systems  \cite{leonov}.  The time average  $
 \bar{f_N}(x) ={1 \over N} \sum^{N-1}_{n=0} f(T^n x) $ behaves as a superposition of quantities which are statistically weakly dependent and fluctuations  around the space average 
$
\langle f \rangle= \int_{M} f(x) d x 
$
have at large $N \rightarrow \infty$ a Gaussian distribution \cite{leonov,Savvidy:2018ygo}:
\be\label{gauss}
\lim_{N\rightarrow \infty}\mu \bigg\{ x	:\sqrt{N}   \left(  \bar{f_N}(x)  - \langle f \rangle \right) < z    \bigg\}
= {1 \over \sqrt{2 \pi} \sigma_f}\int^{z}_{-\infty} e^{-{y^2 \over 2 \sigma^2_f}} dy.
\ee
We were able to express the standard deviation in terms of entropy 
$
\sigma^2_f =\sum^{+ \infty}_{n=-\infty} { M^2  \over   128 \pi^4  } ~  e^{-4 n h(T) }.
$
 It was suggested in 1986 in \cite{yer1986a} to use the hyperbolic C-K systems to generate high quality pseudorandom numbers for Monte-Carlo simulations used in high energy experiments at CERN  for the design of the efficient particle detectors and for the statistical analysis of the experimental data \cite{cern}.  The  entropies of the C-K systems suggested  in \cite{yer1986a,konstantin,Savvidy:2015jva,Savvidy:2015ida,Savvidy:2018ygo,hepforge} are linearly increasing  with the dimension of the operator $T$. The MIXMAX random number generators are currently made available in a portable implementation  in the C++ language at hepforge.org \cite{hepforge} and were implemented into the   Geant4/CLHEP and ROOT toolkits at CERN \cite{cern,root,geant}.

\section{\it Acknowledgement }
This work was presented at the Steklov Mathematical Institute (September 10, 2019) as well as the  CERN Theory Department and A. Alikhanian National Laboratory in Yerevan, where part of this work was completed.  I thank these institutions for their hospitality.   I would like to thank Luis Alvarez-Gaume for stimulating discussions, for kind hospitality at Simons Center for Geometry and Physics and providing to the author the references \cite{hejhal2} and \cite{hejhal}. I would like to thank  H.Babujyan, R.Poghosyan and K.Savvidy for collaboration and enlightening  discussions. The extended presentation is published in arXiv:2001.01785.

\vfill

\end{document}